# Superconductivity in High-Entropy Antimonide $M_{1-x}Pt_xSb$ (M = equimolar Ru, Rh, Pd, and Ir)

Daigorou Hirai*,†, Naoto Uematsu†, Koh Saitoh‡, Naoyuki Katayama† and Koshi Takenaka†

†Department of Applied Physics, Nagoya University, Nagoya 464–8603, Japan

‡Institute of Materials and Systems for Sustainability, Nagoya University, Nagoya 464-8603, Japan

**ABSTRACT:** The high-entropy concept was applied to the synthesis of transition-metal antimonides, $M_{1-x}Pt_xSb$ (M = equimolar Ru, Rh, Pd, and Ir). High-entropy antimonide samples crystallized in a pseudo-hexagonal NiAs-type crystal structure with a $P6_3/mmc$ space group were successfully synthesized through a conventional solid-state reaction and subsequent quenching. A detailed investigation of the composition and equilibration conditions confirmed the reversible phase transition between a multiphase state at low temperature and an entropy-driven single-phase solid solution at high temperatures. Electrical resistivity, magnetization, and heat capacity measurements of single-phase $M_{1-x}Pt_xSb$ ($x$ = 0.2) samples revealed a bulk superconducting transition at 2.15(2) K. This study demonstrates that the high-entropy concept provides numerous opportunities for the discovery of new functional materials such as superconductors.

INTRODUCTION

In recent years, attempts have been made to explore new phases of matter by mixing several elements according to the concept of high-entropy alloys (HEAs)[1,2]. HEAs are typically alloys with five or more constituent elements, each with a molar ratio of 5–35%. A high degree of disorder, that is, a high configurational entropy, is the driving force for material stabilization. The resulting materials exhibit unique mechanical properties that have attracted considerable interest[3].

The thermodynamic stability of a material is given by the Gibbs free energy, $\Delta G = \Delta H - T\Delta S$. Typically, the enthalpy term, $\Delta H$, is the dominant factor, and stable phases are formed by the interactions between the atoms. However, when several elements are mixed at high temperatures, the entropy term, $T\Delta S$, can determine the stability of the material. The ideal entropy of mixing is given by $S = -R\sum x_i \ln(x_i)$, where $x_i$ are the fractions of the constituent elements and $R$ is the gas constant. For an alloy consisting of five equimolar elements, the entropy is equal to $1.61R$. At 1000 K, the reaction temperature of common inorganic solids, the $T\Delta S$ term is equal to 13.37 kJ/mol, which can be as high as the enthalpy of the reactions in solids.

Alloys and compounds containing five or more elements are currently being investigated as high-entropy (HE) materials. However, entropy does not stabilize all these materials. Materials for which entropy has been identified as the decisive term in phase formation are specifically distinguished as "entropy-stabilized" compounds[4,5]. For example, when $\Delta H$ is negative during the synthesis of a compound containing five or more elements, entropy does not play a significant role in phase formation. Conversely, if reaction enthalpy is positive, there exists a critical temperature, $T = \Delta S/\Delta H$, such that $\Delta G = 0 = \Delta H - T\Delta S$, corresponding to the temperature of the transition from the high-temperature phase stabilized by entropy to the low-temperature enthalpy-driven phase (or mixture). This metastable high-temperature phase is generally obtained by quenching to room temperature, at which point the diffusion of solids is sufficiently slow. This entropy-stabilized metastable phase is expected to exhibit a "cocktail effect," that is the appearance of new properties obtained by a complex mixture of elements that cannot be independently obtained from any individual constituent element.

Following the study by Rost *et al.* on entropy-stabilized oxides in 2015[6], the method of entropy-stabilizing new phases has been extended from alloys to other compounds, such as oxides. Consequently, the diversity of the unique properties of HE materials has expanded[7,8], and ionic conductivity, dielectric response, thermoelectric properties, catalysis, thermal barrier coatings, and other novel properties have been reported. In addition to oxides, the range of material systems based on the HEA concept has expanded to include HE materials containing anions other than oxygen, such as carbides[9], nitrides[10], borides[11], and chalcogenides[12]; however, only a limited number of studies have been conducted on these materials.

Since the first HEA superconductor was reported in 2014[13], investigations of superconductivity in HE materials have mainly focused on HEAs. HEA superconductors typically consist primarily of 4d and 5d transition metals (TMs), which mainly form four types of crystal structures: bcc, α-Mn, CsCl, and hcp crystal structures[14]. The bulk nature of the superconductivity in these HEAs has been confirmed by various physical property measurements. One of the striking features of HEA superconductors is their extraordinarily robust superconductivity under pressures of up to ~190 GPa[15]. However, studies on the superconductivity of HE materials are limited.

This study focuses on pnictides. The electronegativities of pnictogens are lower than those of oxygen and are comparable to those of TM elements[16]. The chemical bonds be-tween TMs and pnictogens are intermediates between ionic and covalent bonds. Semiconductors[17], high-performance thermoelectric materials[18], and high-temperature superconductors[19] have been developed to exploit the characteristic chemical bonds of pnictides.

For HE compounds, some studies have been conducted, and the half-Heusler antimonide $MFe_{1-x}Co_xSb$ with six equimolar elements (Ti, Zr, Hf, V, Nb and Ta) on the M site has been reported to be semiconducting and exhibit good thermoelectric performance[20]. However, to the best of our knowledge, entropy stabilization of antimonides has not yet been examined.

**Table 1.** Crystal structure of transition-metal monoantimonides and the ionic radius of the trivalent transition-metal ion[21], electronic configuration of the transition-metal ion, unit cell parameters and volume, and superconducting transition temperature ($T_c$).

| Composition | Structural type | Cation radius (Å) | Electronic configuration | Unit cell parameters (Å) | Cell volume (Å$^3$) | $T_c$ (K) | Ref. |
|---|---|---|---|---|---|---|---|
| RuSb | MnP | 0.68 | 4d$^5$ | $a$ = 5.9608(13)<br>$b$ = 3.7023(9)<br>$c$ = 6.5797(13) | $V$ = 145.21<br>$V/2$ = 72.61 | 1.2 | 22,23 |
| RhSb | MnP | 0.665 | 4d$^6$ | $a$ = 5.9718(7)<br>$b$ = 3.8621(7)<br>$c$ = 6.3242(9) | $V$ = 145.86<br>$V/2$ = 72.93 | — | 23 |
| PdSb | NiAs | 0.76 | 4d$^7$ | $a$ = 4.076(1)<br>$c$ = 5.591(1) | $V$ = 80.44 | 1.32-1.66 | 24–26 |
| IrSb | NiAs | 0.68 | 5d$^6$ | $a$ = 3.978(1)<br>$c$ = 5.521(2) | $V$ = 75.66 | — | 27 |
| PtSb | NiAs | 0.625 (4+) | 5d$^7$ | $a$ = 4.1318(6)<br>$c$ = 5.483(1) | $V$ = 81.06 | 2.1 | 28,29 |
| $M_{1-x}Pt_xSb$ $x$ = 0.2 | NiAs | 0.682* | d$^{6.2}$* | $a$ = 4.0022(1)<br>$c$ = 5.56719(9) | $V$ = 77.227(3) | 2.15(2) | This work |

*Average cation radii and d-electron counts

As shown in Table 1, two closely related crystal structure types are found in TM monoantimonides (TMSb): a NiAs-type structure with a $P6_3/mmc$ (#194) space group and an MnP-type structure with a $Pnma$ (#62) space group (Fig. 1a).

In both structures, the TM atom coordinated with six Sb atoms to form a $TMSb_6$ octahedron (Fig. 1a). In the NiAs-type structure, the distances between the TM and its coordinating Sb atoms were equal, and the TM atoms formed an equilateral triangular lattice on the (001) plane. In contrast, in the MnP-type structure, the $TMSb_6$ octahedron was significantly distorted because of the formation of Sb-Sb and TM-TM covalent bonds[30]. According to the previous studies, TMSb crystallizes in the MnP-type structure of Ru and Rh at the TM site and in the NiAs-type structure of Pd, Ir, and Pt (Table 1) [23,25,27].

In this study, a single-phase solid solution of $M_{1-x}Pt_xSb$ (M = equimolar amounts of Ru, Rh, Pd, and Ir) was synthesized according to the HE concept. The entropy-driven stabilization of the solid solution with the NiAs-type structure was confirmed by varying the composition and equilibration conditions. Single-phase samples of $M_{1-x}Pt_xSb$ with $x$ = 0.2 exhibited a bulk superconducting transition at $T_c$ of 2.15(2), which was comparable to the highest $T_c$ in single-TM-element TMSb materials.

EXPERIMENTAL SECTION

All $M_{1-x}Pt_xSb$ (M = equimolar Ru, Rh, Pd, and Ir) samples were prepared using the conventional solid-state reaction. Stoichiometric amounts of elemental Sb (99.99%, Rare Metallic Co., Ltd), Ru, Rh (99.95%, Rare Metallic Co., Ltd), Pd, Ir (99.9%, Kojundo Chemical Laboratory Co., Ltd), and Pt (99.95%, Tanaka Kikinzoku Kogyo K.K.) were thoroughly homogenized using an agate mortar and pestle. The obtained powder was pressed into a pellet with a diameter of 6 mm using a hand press. The pellet was sealed in an evacuated quartz tube, heated to 1000 °C at a rate of 200 °C/h, maintained for 24 h, and then furnace-cooled.

For the phase stability study, the pelletized samples were heated to the target temperature in an evacuated quartz tube at a rate of 200 K/h, equilibrated at that temperature, and finally air-quenched to room temperature.

The obtained samples were characterized using X-ray diffraction (XRD) at room temperature with Cu-K$\alpha$ radiation using a diffractometer (Rigaku, Minflex). For a detailed structural analysis, synchrotron powder XRD experiments were conducted at 300 K and an X-ray energy of 20 keV using a quadruple PILATUS 100K detector at the BL5S2 of the Aichi Synchrotron Radiation Center. Powdered samples were sealed in a Lindemann glass capillary of 0.1 mm diameter. RIETAN-FP was used for Rietveld analysis[31].

High-resolution scanning transmission electron microscopy (STEM) was performed using an aberration corrected scanning transmission electron microscope equipped with a cold field-emission gun (JEOL, ARM200F-Cold-FE) operated at an acceleration voltage of 200 kV. The $M_{1-x}Pt_xSb$ sample was crushed and dispersed onto a holy amorphous carbon film on a copper grid for electron microscopy. The convergence semi angle of the incident electron beam was set to 15 mrad. An annular dark-field (ADF) detector was used to collect the electrons scattered over an angular range of 68–280 mrad. Under these conditions, the image contrast had a strong atomic number ($Z$) dependence that was proportional to almost the square of $Z$. Atomic resolution elemental mapping was performed using energy-dispersive X-ray spectroscopy (EDS) system (JEOL, JDE-2300T and Thermo Scientific, Noran system seven) equipped with an aberration-corrected STEM instrument. Approximately 250 spectrum images, taken with a pixel size of 128 × 128 at a pixel time of 10 μs for each frame, were drift-corrected and integrated. Quantitative EDS analysis was performed using the Cliff–Lorimer method with absorption correction.

Electrical resistivity and heat capacity measurements were performed using a physical property measurement system (PPMS, Quantum Design). Electrical resistivity was measured using the four-probe method with silver paste (DuPont 4922N, DuPont) terminals. The heat capacity was measured using a relaxation method. A thin pellet sample weighting approximately 20 mg was mounted on a sapphire sample holder using grease (Apiezon-N, M&I Materials) to ensure good thermal contact. The magnetization was measured using a magnetic property measurement system (MPMS3, Quantum Design). A piece of the pelletized sample (approximately 20 mg) was mounted on a quartz rod with varnish (GE7031, General Electric Company).

RESULTS AND DISCUSSION

The results for the thermodynamic stability of the phases are first discussed, followed by their physical properties, including superconductivity.

Figures 1b-e show the XRD patterns of a $M_{1-x}Pt_xSb$ (M = equimolar Ru, Rh, Pd, and Ir) sample with $x$ = 0.2 that contained all the TMs in equimolar amounts and had the maximum configurational entropy after heat treatment under various conditions.

First, the raw materials were mixed, pelletized, pre-reacted at 1000 °C for 24 h, and furnace cooled. The sample was then crushed, pelletized, sealed again in an evacuated quartz tube, and equilibrated at 800 °C for 24 h. The quartz tube was quenched in air at room temperature. All the major peaks in the XRD pattern were indexed to the NiAs-type structure; however, a few peaks were also observed, which were attributed to the MnP-type impurity phase (Fig. 1b). The biphasic sample was pelleted, equilibrated at 1100 °C in a vacuum quartz tube, and quenched to room temperature. The sample equilibrated at 1100 °C was a single NiAs-type solid solution without any impurity phases (Fig. 1c). After further equilibration at 800 °C and following quenching, the MnP-type impurity phase again precipitated from the host NiAs-type phase (Fig. 1d), and equilibration at 1100 °C again recovered a single-phase state (Fig. 1e). Therefore, the two-phase separation at 800 °C and single-phase state at 1100 °C are thermodynamically stable states, and the two states can be reversibly transformed into each other across a certain phase transition temperature. Further investigation of the equilibration conditions revealed that an enantiotropic (reversible with respect to temperature) phase transition occurred between 800 and 900 °C.

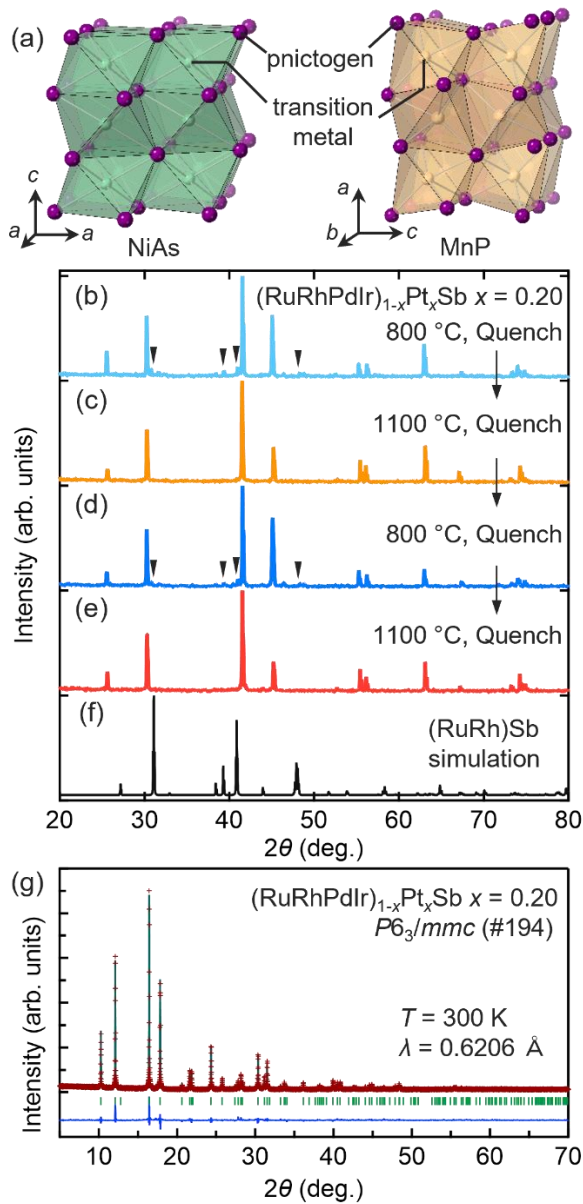

**Figure 1.** (a) Comparison of NiAs- and MnP-type crystal structures. (b-e) XRD patterns of a powdered sample of $M_{1-x}Pt_xSb$ (M = equimolar Ru, Rh, Pd, Ir) with $x = 0.2$ after sequential heat treatment in an evacuated quartz tube. (b) First, the sample reacted at 1000 °C for 24 h was ground and repelletized. The pellet was equilibrated for 24 h at 800 °C, followed by air-quenching to room temperature (RT). (c) Second, the same sample was equilibrated for 24 h at 1100 °C and quenched to RT. (d) Third, the sample was equilibrated for 24 h at 800 °C followed by quenching. (e) Finally, equilibration for 24 h at 1100 °C followed by quenching was performed. All samples contain NiAs-type phase as the main phase. Triangles in (b) and (d) indicate the MnP-type impurity phase. The simulation pattern for (RuRh)Sb with MnP-type structure is shown in (f). (g) Synchrotron XRD pattern of a powdered sample of $M_{1-x}Pt_xSb$ with $x = 0.2$ quenched from 1100 °C measured at 300 K and Rietveld fitting. Observed (red crosses), calculated (black line), and difference (lower blue line) XRD patterns are displayed. Green tick marks indicate the positions of allowed reflections.

This behavior is consistent with the entropy-driven stabilization of the NiAs-type solid solution at high temperatures, in which the NiAs-type phase is unstable when enthalpy alone is considered, resulting in a two-phase separation at low temperatures. However, the entropy term overcomes the positive enthalpy of mixing at high temperatures, and the NiAs-type phase with a disordered arrangement of TM atoms becomes stable. The quantitative evaluation of the transition temperatures described below further supports the entropy-driven stabilization.

For detailed structural analysis, Rietveld refinement was performed to a synchrotron XRD pattern at 300 K for a powdered sample of $M_{1-x}Pt_xSb$ with $x = 0.2$ quenched from 1100 °C. The sharp peaks in the XRD pattern indicate the high crystallinity of the sample. The good agreement between the experimental data and simulated pattern indicates the validity of the obtained structural parameters (Fig. 1g). The crystallographic and Rietveld refinement data are summarized in Supporting Information, Table S1.

The lattice constants of the single-phase $M_{1-x}Pt_xSb$ ($x = 0.2$) sample were $a = 4.0022(1)$ Å and $c = 5.56719(9)$ Å, and the cell volume was $V = 77.227(3)$ Å$^3$. The unit cell volume of the MnP-type structure tended to be smaller than that of the NiAs-type structure. Comparing Sb compounds containing Ru and Ir ions, which possess the same ionic radii, the volume of MnP-type RuSb ($V = 72.61$ Å$^3$) was smaller than that of NiAs-type IrSb ($V = 75.66$ Å$^3$). The large unit cell volume of $M_{1-x}Pt_xSb$ was attributed to its NiAs-type structure, in which Ru and Rh ions could not naturally crystallize.

The Debye–Waller factor (DWF, $B$) for the transition-metal site was calculated to be 1.87(3) Å$^2$, which is a relatively large value for inorganic compounds. On the contrary, similarly large DWFs have been reported for other HE materials: $B = 1.3(3)$ Å$^2$ for (Ti, Zr, Hf, Sn)O$_2$[32] and $B = 2.5$ Å$^2$ for (ScZrNb)$_{0.6}$(RhPd)$_{0.4}$[33]. HE materials tend to exhibit large DWF, probably because of the random distribution of their constituent elements. The large DWF of $M_{1-x}Pt_xSb$ ($x = 0.2$) seemed to reflect a high degree of disorder in the occupation of the transition-metal atoms.

To examine atomic-scale homogeneity, structural and chemical analyses were conducted using high-resolution high-angle annular dark-field (HAADF) STEM, convergent-beam electron diffraction (CBED), and EDS, as shown in Fig. 2.

Six-fold rotational symmetry was clearly observed in the HAADF-STEM image and the CBED pattern recorded along the [0001] zone axis for a quenched powdered sample of $M_{1-x}Pt_xSb$ with $x = 0.2$ (Figs. 2a and b). This was consistent with the space group $P6_3/mmc$. The larger bright and smaller dark spots in Fig. 2a corresponded to the transition-metal and antimony atoms, respectively. The regular alignment of the atoms confirmed the pseudo-hexagonal NiAs-type structure, even at the nanoscale. The high crystallinity of the sample was evidenced by sharp reflection spots in the selected-area electron diffraction pattern (Supporting Information, Fig. S1a).

As shown in Fig. 2c, all the constituent elements were uniformly distributed in the atomic-scale EDS maps, demonstrating local chemical homogeneity. The quantitative analysis of the EDS spectra obtained from this region showed that the compositions of Ru, Rh, Pd, Ir, Pt, and Sb in atomic percentages were 10.8, 12.5, 9.4, 9.3, 8.4, and 49.6, respectively, with an error of approximately 1%. The obtained composition was consistent with the nominal value of Ru : Rh : Pd : Ir : Pt : Sb = 10 : 10 : 10 : 10 : 10 : 50. The results of compositional analyses for other samples between $x = 0.1$ and $x = 0.4$ (Supporting Information, Table S2) confirmed that the actual atomic ratio was close to the nominal value.

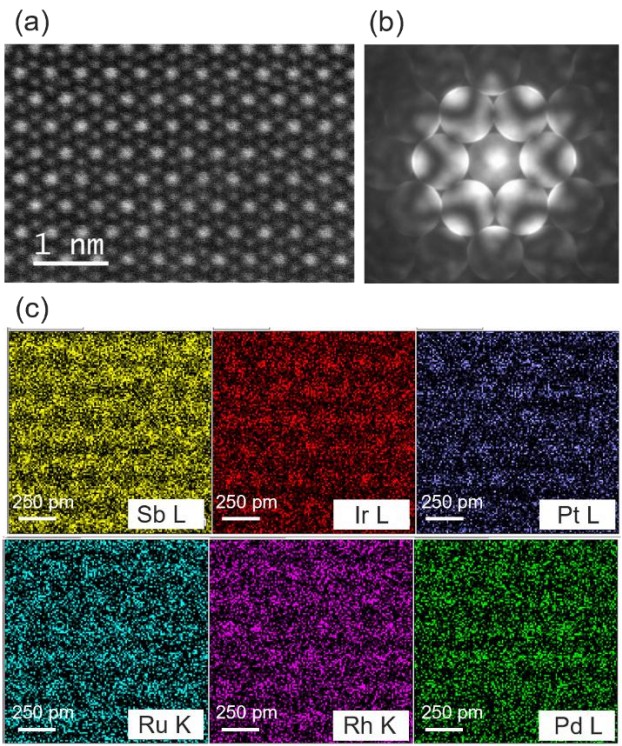

**Figure 2.** (a) High-resolution high-angle annular dark-field scanning transmission electron microscopy (HAADF-STEM) image and (b) convergent-beam electron diffraction (CBED) pattern recorded along the [0001] zone axis and (c) energy dispersive X-ray spectroscopy (EDS) maps for all the elements present in a quenched powdered sample of $M_{1-x}Pt_xSb$ (M = equimolar Ru, Rh, Pd, Ir) with $x = 0.2$.

To clarify the compositional dependence of the phase stability, XRD patterns were compared for the quenched samples of $M_{1-x}Pt_xSb$ with varying $x$, where all samples were equilibrated at 1100 °C for 24 h (Fig. 3). For $x = 0$, which did not contain Pt and contained 0.25 moles of Ru, Rh, Pd, and Ir, a mixture of NiAs- and MnP-type phases were observed. However, for $x \geq 0.05$, all the samples were single-phase solid solutions with a NiAs-type structure. The unit cell parameters obtained by parameter fitting showed a systematic dependence on $x$ (Supporting Information Fig. S2).

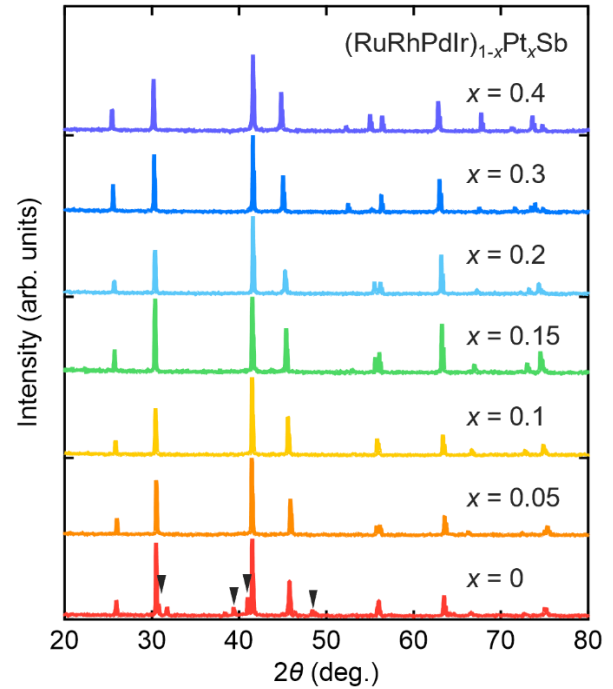

**Figure 3.** XRD patterns of powdered samples of $M_{1-x}Pt_xSb$ (M = equimolar Ru, Rh, Pd, and Ir) with $x$ ranging from 0 to 0.4 after equilibration for 12 h at 1100 °C followed by air quenching to room temperature. For all the compositions, the main phase crystallizes in the NiAs-type structure. Triangles indicate XRD peaks from the MnP-type phase.

To determine the compositional dependence of the phase transition temperature between the single-phase and multiphase states, equilibration at various temperatures, followed by quenching, was performed for each composition. The obtained composition-temperature phase diagram is shown in Fig. 4a. For all the compositions, a single NiAs-type phase was obtained at high temperatures, and below a certain phase transition temperature, $T_{mix}$, a multiphase state appeared. $T_{mix}$ was defined as the midpoint between the lowest temperature at which a single phase was obtained, and the highest temperature at which a multiphase state occurred.

Phase stability was determined using the Gibbs free energy. The phase transition temperature $T_{mix}$ is the point at which the difference in the Gibbs free energy between the multiphase and single-phase states ($\Delta G_{mix}$) is zero, so that $\Delta G_{mix} = 0 = \Delta H_{mix} - T_{mix}\Delta S_{mix}$ and the transition temperature is expressed as $T_{mix} = \Delta H_{mix}/\Delta S_{mix}$, where $\Delta H_{mix}$ is the enthalpy cost required to achieve the single-phase state and $\Delta S_{mix}$ is the entropy difference between the two states. $\Delta H_{mix}$ was positive in $M_{1-x}Pt_xSb$ because the entropy term decreases at low temperatures, resulting in a two-phase separation.

Simplifying $\Delta S_{mix}$ as the ideal configurational entropy yielded an upper convex curve with a maximum value at $x$ = 0.2, where all the TMs were equimolar, as shown by the solid line in Fig. 4b. Given that $\Delta H_{mix}$ was independent of the composition, $T_{mix}$ was proportional to $1/\Delta S_{mix}$ and thus had a lower convex composition dependence, with a minimum at $x$ = 0.2. For the entropy-stabilized oxide (MgCoNiCuZn)O, the compositional dependence of the transition temperature followed a lower convex curve that reached a minimum value at an equimolar composition, implying an entropy-driven transition[6].

However, if $\Delta H_{mix}$ depended on the composition, the compositional dependence should be strongly reflected in $T_{mix}$, and the monotonic decrease in $T_{mix}$ with increasing $x$ in $M_{1-x}Pt_xSb$ suggested that $\Delta H_{mix}$ also decreased with increasing $x$. Because the single-phase state at high temperatures had a NiAs-type structure, the dominant contribution of $\Delta H_{mix}$ was considered to be the transition energy, $\Delta H_{NiAs-MnP}$, of RuSb and RhSb from their regular MnP-type structures to NiAs-type forms. Under this assumption, $\Delta H_{mix}$ was proportional to the combined Ru and Rh molar fractions, $M_{Ru,Rh}$, which was 0.5 at $x$ = 0 and decreased monotonically with increasing $x$ toward $M_{Ru,Rh}$ = 0.3, and $x$ = 0.4. Therefore, $\Delta H_{mix}$ decreased monotonically with an increase $x$.

From the transition temperature $T_{mix}$ = 1123 K at $x$ = 0.2, $\Delta H_{mix}$ = 15.0 kJ was obtained assuming the entropy difference $\Delta S_{mix}$ = 1.61$R$. If the $\Delta H_{mix}$ was provided by Ru and Rh atoms, using $M_{Ru,Rh}$ = 0.4 at $x$ = 0.2, $\Delta H_{NiAs-MnP}$ was estimated to be 37.6 kJ mol$^{-1}$. The chemical potential changes for the wurtzite-to-rocksalt and tenorite-to-rocksalt transitions of ZnO and CuO have been reported to be 25 and 22 kJ mol$^{-1}$[6], respectively. During these transitions, the coordination numbers of the TM atoms changed. The obtained value of 37.6 kJ mol$^{-1}$ for $\Delta H_{NiAs-MnP}$ was unreasonably large, because the change in chemical potential from the MnP-type structure to the NiAs-type structure, where the coordination number remained constant, was expected to be smaller.

The overestimation of $\Delta H_{mix}$ was likely due to the inaccuracy in the estimation of $\Delta S_{mix}$. Under the above assumption, $\Delta S_{mix}$ was the difference in configurational entropy between the single-phase solid solution and the multiphase state consisting of five phases with different crystal structures. However, the phase-separated state at low temperatures was a mixture of a MnP-type phase with a disordered arrangement of Ru and Rh and the NiAs-type phase with a disordered arrangement of Pd, Pt, and Ir. Even in the phase-separated state at low temperatures, the entropy of the configuration was as high as 0.94$R$, so the entropy difference from the single-phase state was only $\Delta S_{mix}$ = 0.67$R$. A reevaluation of $\Delta H_{NiAs-MnP}$ obtained 15.7 kJ mol$^{-1}$ which was reasonable.

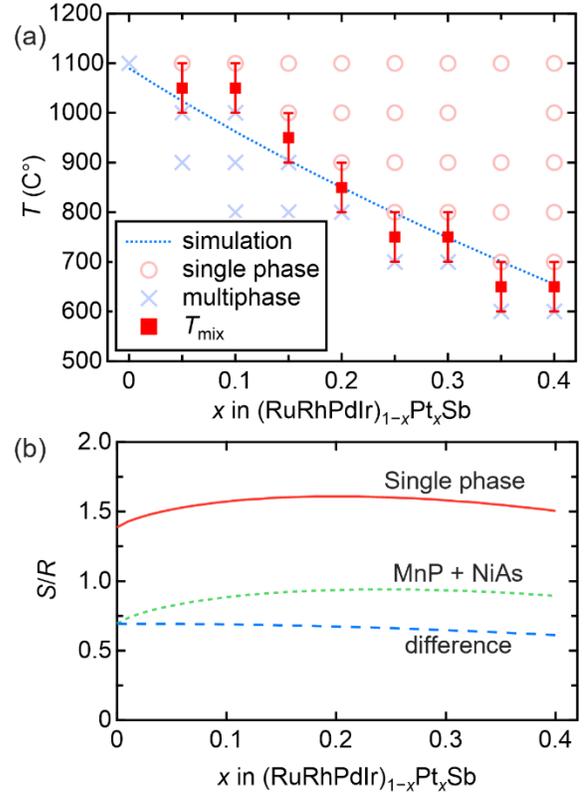

**Figure 4.** (a) Temperature-composition phase diagram for $M_{1-x}Pt_xSb$ (M = equimolar Ru, Rh, Pd, and Ir) showing the transition temperature, $T_{mix}$ (red squares), between the high-temperature single-phase and low-temperature multiphase states as a function of composition in the Pt content range from 0 to 0.4. Light red open circles and light blue crosses represent single-phase and multiphase states, respectively. Blue dotted line denotes the simulated $T_{mix}$, obtained from the Gibbs free energy. (b) Calculated configurational entropy as a function of $x$ for the high-temperature single-phase state (red solid line), the low-temperature mixture of MnP- and NiAs-type phases (green dotted line), and the difference between them (blue broken line).

The configurational entropies of the high- and low-temperature phases and their differences are indicated by the red solid, green dotted, and blue broken lines in Fig. 4b, respectively. Under the assumptions that $\Delta H_{NiAs-MnP}$ = 15.7 kJ mol$^{-1}$ and $\Delta H_{mix} = \Delta H_{MnP-NiAs} \times M_{Ru,Rh}$, the compositional dependence of $T_{mix}$ was calculated (blue dotted line in Fig. 4a). The simulated curve reasonably reproduced the actual transition temperatures, demonstrating the validity of this simplified thermodynamic model. According to this consideration, any element that constitutes antimony compounds with NiAs type structure, can be replaced with Pd, Pt, and Ir to synthesize HE antimonides with a NiAs-type structure without significant enthalpy cost. Thus, Ti, V, Cr, Mn, Fe, Co, Ni, and Nb are potential elements for inclusion in HE antimonides.

To reveal the properties of the HE antimonides, physical properties were measured for $M_{1-x}Pt_xSb$ samples with $x$ = 0, 0.1, 0.2, 0.3, and 0.4 quenched at 1100 °C. The temperature dependences of electrical resistivity and magnetic susceptibility are shown in Fig. 5. Except for $x$ = 0.1, the resistivities at 300 K were low (less than 0.2 mΩ cm), indicating metallic conductivity. However, the resistivity exhibited a considerably weak temperature dependence, with a gradual increase toward lower temperatures; this behavior is quite different from that of ordinary metals. The exceptionally large resistivity at $x$ = 0.1 can be attributed to the fact that $x$ = 0.1 was quenched near the phase-transition temperature and thus had a stronger disorder than the other compositions. Furthermore, at low temperatures, the electrical resistivity of $x$ = 0.2 sample abruptly dropped at 2.3 K, and became zero, suggesting a superconducting transition (Figs. 5a and 6a).

In most HEAs, the electrical resistivity decreases with decreasing temperature, although the temperature dependence is weak. An exception is $Al_{2.08}CoCrFeNi$, for which near-constant resistivity was observed over a wide temperature range from 360 to 4.2 K.[34] This behavior was attributed to the strong scattering of conduction electrons caused by the highly disordered arrangement of metal atoms. In many metals, including A15 and heavy fermion superconductors, the electrical resistivity saturates under the Mott–Ioffe–Regel condition, in which the mean free path of the electrons is comparable to the interatomic distances[35–37]. Similarly, in $M_{1-x}Pt_xSb$, the conduction electrons are likely to be strongly scattered owing to the high degree of disorder, as suggested by the large DWF. It would be interesting if electron scattering was enhanced not only by static disorder but also by other interactions such as electron-electron and electron-phonon interactions.

The magnetic susceptibilities ($\chi$s) of all the samples had small negative values (approximately $-4 \times 10^{-5}$ cm$^3$ mol$^{-1}$ at 300 K) that was almost independent of temperature. Considering $M_{1-x}Pt_xSb$ as a metal, the observed $\chi$ was likely a combination of small Pauli paramagnetic and core diamagnetic susceptibilities.

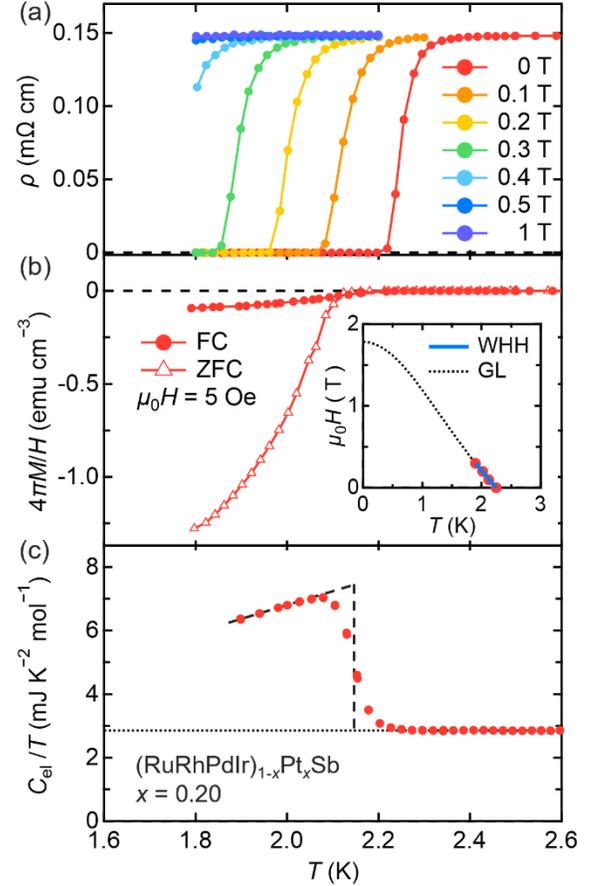

**Figure 6.** Temperature dependences of (a) electrical resistivity in magnetic fields from 0 to 1 T, (b) magnetization, and (c) the electronic contribution of the heat capacity divided by the temperature displaying the superconducting transition in $M_{1-x}Pt_xSb$ (M = equimolar Ru, Rh, Pd, and Ir) with $x$ =0.2. Magnetization was measured in zero-field cooling (ZFC) and field cooling (FC) processes in a field of 5 Oe. Dotted line in (c) represents the Sommerfeld constant $\gamma$. The $T_c$ was determined by considering entropy conservation near $T_c$, as indicated by the broken line in (c). Inset in (b) shows the temperature dependence of the upper critical field, $H_{c2}(T)$, estimated from the midpoint of the superconducting transition in the electrical resistivity. Werthamer–Helfand–Hohenberg (WHH, blue solid line) and Ginzburg–Landau (GL, black dotted line) fittings are shown.

Using low-temperature magnetization measurements, the phase transition at 2.3 K in $x$ = 0.2 sample was confirmed to be a superconducting transition (Fig. 6b). A zero-field-cooled (ZFC) scan revealed a large diamagnetic signal below 2.1 K, indicating superconductivity, reaching 128% of perfect diamagnetism at 1.8 K. The bulk nature of the superconducting transition was further confirmed by heat capacity measurements (Fig. 6c). A clear jump, indicating a bulk phase transition was observed in the electronic contribution of the heat capacity divided by the temperature $C_e/T(T)$, at the same temperature at which zero resistance and large diamagnetism occurred. The jump in $C_e/T(T)$ was sharp, demonstrating that the sample homogeneously transformed into a superconducting state despite its highly disordered cation configuration. Considering the entropy balance near the transition, the transition temperature was determined as $T_c$ = 2.15(2) K.

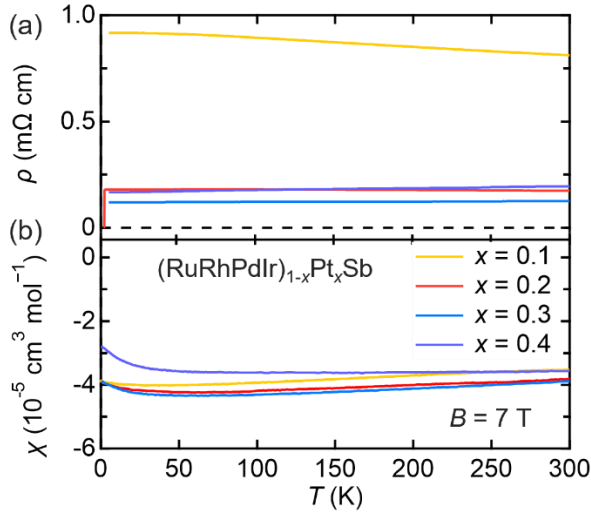

**Figure 5.** Temperature dependences of (a) electrical resistivity and (b) magnetic susceptibility in a magnetic field of 7 T for $M_{1-x}Pt_xSb$ (M = equimolar Ru, Rh, Pd, and Ir) with $x$ = 0.1, 0.2, 0.3, and 0.4.

The heat capacity of the normal state just above $T_c$ was reproduced well by $C/T(T) = \gamma + \beta T^2$, yielding $\gamma$ = 2.85(2) mJ mol$^{-1}$ K$^{-2}$ and $\beta$ = 0.346(3) mJ mol$^{-1}$ K$^{-4}$ by fitting the data. A Debye temperature ($\Theta_D$) of 224 K was obtained from the $\beta$ value[38]; the obtained $\Theta_D$ was comparable to those of other TM antimonides: $\Theta_D$ = 212 K for RuSb[39], $\Theta_D$ = 308 K for IrSb$_3$[40], and $\Theta_D$ = 250 K for MgPd$_2$Sb[41]. The comparable $\Theta_D$ values indicate that neither significant softening nor hardening of the lattice vibrations occurred in the HE antimonides. The Sommerfeld constant $\gamma$ was smaller than in the analogous antimony compounds: $\gamma$ = 5.5 mJ mol$^{-1}$ K$^{-2}$ for RuSb[39] $\gamma$ = 2.63 mJ mol$^{-1}$ K$^{-2}$ for PdSb[26], and $\gamma$ = 4.9 mJ mol$^{-1}$ K$^{-2}$ for MgPd$_2$Sb[41]. The smaller $\gamma$ values were likely due to the high degree of disorder at the TM sites, which reduced in the density of states (DOS) at the Fermi energy ($E_F$). Using the obtained Sommerfeld constant to evaluate the jump in heat capacity, $\Delta C/(\gamma T_c)$ = 1.56 was obtained, which was slightly larger than the value of 1.43 expected for conventional weak-coupling BCS superconductors[42]. The first HEA superconductor, Ta$_{34}$Nb$_{33}$Hf$_8$Zr$_{14}$Ti$_{11}$, also showed a heat capacity jump of $\Delta C/(\gamma T_c)$ = 1.63[13]. Analogous HEAs have also been reported with heat capacity jumps at $T_c$ exceeding 1.43[43]. It is desirable to conduct a comprehensive study of the superconducting coupling strengths of various HE materials to discover the common characteristics of superconductivity in these materials.

To examine the superconducting properties, electrical resistivities were measured under various magnetic fields. As shown in Fig. 6a, $T_c$ decreased with increasing applied magnetic field and was finally suppressed below 1.8 K in a magnetic field of 1 T. The temperature dependence of the upper critical field $\mu_0 H_{c2}(T)$ obtained from these results is shown in the inset of Fig. 6b. $T_c$ is defined as the midpoint of the superconducting transition. Using Werthamer–Helfand–Hohenberg (WHH) formula[44]: $\mu_0 H_{c2}(0) = 0.69 \, T_c |d\mu_0 H_{c2}/dT|_{T=T_c}$, $\mu_0 H_{c2}(0)^{WHH}$ was calculated to be 1.33 T from the slope close to the $T_c$. Fitting the data to the Ginzburg–Landau (GL) equation ($\mu_0 H_{c2}(T) = \mu_0 H_{c2}(0) \times \{1 - (T/T_c)^2\}/\{1 + (T/T_c)^2\}$) yielded $\mu_0 H_{c2}(0)^{GL}$ = 1.78 T. Due to the limited temperature range of the measurement, these values may contain large errors.

The observed upper critical fields were relatively high but did not exceed the Pauli paramagnetic limit of 3.96 T. A coherence length $\xi_0$ of 160 Å was obtained from $H_{c2}(0)^{GL}$, which is lower than that of other superconductors with similar $T_c$ values: $T_c$ = 1.32 K, $\mu_0 H_{c2}(0)$ = 0.09 T for PdSb[26], $T_c$ = 2.2 K, $\mu_0 H_{c2}(0)$ = 0.194 T、$\xi_0$ = 412 Å for MgPd$_2$Sb[41], $T_c$ = 2.1 K, $\mu_0 H_{c2}(0)$ = 0.14 T, $\xi_0$ = 480 Å for BaIr$_2$P$_2$[45,46]. $H_{c2}(0)$ tends to be relatively high in HEA superconductors[14,43]. The highly disordered arrangement of the TM atoms may be related to their short coherence lengths.

Since the first HEA superconductor was reported in 2014[13], investigations of superconductivity in HE materials have mainly focused on HEAs[14]. Several HEA superconductors with different compositions and crystal structures have been developed and attempts have been made to understand them comprehensively. For HEA superconductors, it was found that the valence electron count (VEC) had a significant effect not only on the stability of HEAs but also on $T_c$s[14,43]. HEA superconductors with a bcc structure exhibited the highest $T_c$ of 7.3 K when the VEC was 4.7, a feature similar to the Matthias plot of $T_c$ versus VEC for a simple binary TM alloy[47]. A comparison of the HEA $T_c$ to those of crystalline binary TM alloys and amorphous thin-film metals showed that the $T_c$ of HEA was lower than that of crystalline alloys but higher than that of amorphous alloys. This indicated that for the same VEC, crystallinity was strongly related to $T_c$. This disorder disrupted the electron dispersion relationship and reduced the DOS at $E_F$, which may be responsible for the lower $T_c$.

Among TMSb, superconducting transitions have been reported at $T_c$ = 1.2, 1.32, and 2.1 K for TM = Ru, Pd, and Pt, respectively (Table 1). The newly discovered M$_{1-x}$Pt$_x$Sb with $x$ = 0.2 has $T_c$ = 2.15(2) K which is comparable to the highest $T_c$ among TMSb examined to date. Given that the superconductivity in M$_{1-x}$Pt$_x$Sb is mediated by the conventional electron-phonon mechanism, the origin of the high $T_c$ is discussed. First, the DOS of M$_{1-x}$Pt$_x$Sb at $E_F$ is expected to be smaller than that of the other TMSb because of its lower Sommerfeld constant than that of RuSb, which has a $T_c$ approximately half that of M$_{1-x}$Pt$_x$Sb $T_c$ ($T_c$ = 1.2 K). This may be due to disorder, as observed in HEA superconductors. Because the Debye temperature is almost the same as that of the other antimonides, the lattice energies contributing to the superconductivity are also comparable. Therefore, one remaining possibility is the enhancement of the electron-phonon coupling. Indeed, the jump in the specific heat at $T_c$ is slightly larger than that expected for the BCS weak-coupling limit.

In the HE compound (CoAu)$_{0.2}$(RhIrPdPt)$_{0.8}$Te$_2$, a $T_c$ of 3.6 K has been observed, which is comparable to the highest $T_c$ reported for metal-ditelluride system[48]. It has been argued that structural instability may contribute to a higher $T_c$. Interestingly, entropy stabilization plays a role in increasing $T_c$. Understanding the microscopic physics of superconductivity in the HE materials is an exciting and challenging task.

CONCLUSIONS

In this study, M$_{1-x}$Pt$_x$Sb (M = equimolar Ru, Rh, Pd, and Ir) was synthesized with various Pt contents, $x$, and heat treatment conditions to obtain entropy-stabilized transition metal monoantimonides. An entropy-stabilized single-phase solid solution with a NiAs-type structure was obtained by equilibration at high temperatures followed by quenching. The strong $x$ dependence of the phase transition temperature was attributed to the composition dependence of the enthalpy. A bulk superconducting transition at $T_c$ = 2.15(2) was observed in the sample with $x$ = 0.2. This $T_c$ is as high as the highest $T_c$ previously observed for TMSb, indicating that the HE concept is an effective means of discovering novel superconductors and improving $T_c$.

## ASSOCIATED CONTENT

**Supporting Information**.
Supporting Information Available:
Crystallographic data (CIF).
Results of Rietveld refinement, chemical analysis, and compositional dependence of unit cell parameters.

## AUTHOR INFORMATION

**Corresponding Author**

*dhirai@nuap.nagoya-u.ac.jp

**Author Contributions**

The manuscript was written with contributions from all authors. All the authors approved the final version of the manuscript.

**Funding Sources**

This work was partly supported by the Japan Society for the Promotion of Science (JSPS) KAKENHI Grant Numbers JP20H01858 and JP22H04462 (Quantum Liquid Crystals).

**Notes**
The authors declare no competing financial interest.

## ACKNOWLEDGMENT

This work was partly conducted under the Visiting Researcher's Program of the Institute for Solid State Physics, the University of Tokyo. Synchrotron powder XRD experiments were conducted at the BL5S2 of the Aichi Synchrotron Radiation Center, Aichi Science and Technology Foundation, Aichi, Japan (Proposal No. 202206142).

**Supporting Information**

## a. Rietveld refinement for $M_{1-x}Pt_xSb$ (M = equimolar Ru, Rh, Pd, Ir) with $x = 0.2$

For detailed structural analysis, synchrotron powder XRD experiments were conducted at 300 K and an X-ray energy of 20 keV using a quadruple PILATUS 100K detector at the BL5S2 of the Aichi Synchrotron Radiation Center. A powder sample of $M_{1-x}Pt_xSb$ (M = equimolar Ru, Rh, Pd, Ir) with $x = 0.2$ quenched from 1100 °C was sealed it in a Lindemann glass capillary of 0.1 mm diameter. RIETAN-FP was used for the Rietveld analysis [1]. The obtained crystallographic and Rietveld refinement data are listed in Table S1.

Table S1. Crystallographic and Rietveld refinement data for a powder sample of $M_{1-x}Pt_xSb$ (M = equimolar Ru, Rh, Pd, Ir) with $x = 0.2$ equilibrated for 24 h at 1100 °C. The equivalent isotropic atomic displacement parameter $B$ and the occupancy for all the transition metals were assumed to be the same.

| Chemical formula: $Ru_{0.2}Rh_{0.2}Pd_{0.2}Ir_{0.2}Pt_{0.2}Sb$ | | | | | | |
|---|---|---|---|---|---|---|
| Temperature: 300 K | | | | | | |
| Wavelength: 0.6206 Å | | | | | | |
| space group: $P6_3/mmc$ (No. 194) | | | | | | |
| $a = b = 4.0022(1)$ Å, $c = 5.56719(9)$ Å | | | | | | |
| $\alpha = \beta = 90°$, $\gamma = 120°$, $V = 77.227(3)$ Å$^3$ | | | | | | |
| Site | Wyckoff | $x$ | $y$ | $z$ | Occupancy | $B$(Å$^2$) |
| Ru1 | $2a$ | 0 | 0 | 0 | 0.2 | 1.87(3) |
| Rh1 | $2a$ | 0 | 0 | 0 | 0.2 | 1.87 |
| Pd1 | $2a$ | 0 | 0 | 0 | 0.2 | 1.87 |
| Ir1 | $2a$ | 0 | 0 | 0 | 0.2 | 1.87 |
| Pt1 | $2a$ | 0 | 0 | 0 | 0.2 | 1.87 |
| Sb1 | $2c$ | 1/3 | 2/3 | 1/4 | 1 | 1.71(3) |
| Rietveld agreement factors | | | | | | |
| $R_{wp} = 8.543$ %, $R_p = 6.872$ %, $R_e = 6.175$ %, $S = 1.3833$ | | | | | | |

### b. Structural and chemical analysis via STEM-EDS

High-resolution scanning transmission electron microscopy (STEM) was conducted by an aberration corrected scanning transmission electron microscope with a cold field emission gun (JEOL, ARM200F-Cold-FE) operated at an acceleration voltage of 200 kV. The $M_{1-x}Pt_xSb$ sample was crushed and dispersed on holy amorphous carbon film on cupper grids for electron microscopy. The convergence semi-angle of the incident electron beam was set to be 15 mrad. Annular dark-field (ADF) detector was set to collect electrons scattered at an angular range of 68 - 280 mrads. Atomic resolution elemental mapping was performed by using energy-dispersive X-ray spectroscopy (EDS) system (JEOL, JDE-2300T and Thermo Scientific, Noran system seven) equipped with the aberration corrected STEM instrument. Approximately 250 spectrum-images taken with a pixel size of 128 x 128 at a pixel time of 10 µsec for each frame were drift-corrected and integrated. Quantitative EDS analysis was performed by the Cliff-Lorimer method with an absorption correction. The obtained EDS spectra are shown in Fig, S1b and the results of chemical analysis via STEM-EDS are summarized in Table S2. The sharp diffraction spots in selected-area electron diffraction pattern demonstrate the high crystallinity of the sample (Fig. S1a).

Fig. S1. (a) Selected-area electron diffraction pattern taken along the [0001] zone axis and (b) energy dispersive X-ray spectroscopy (EDS) spectra for a quenched powder sample of $M_{1-x}Pt_xSb$ (M = equimolar Ru, Rh, Pd, Ir) with $x = 0.2$.

Table S2. Results of chemical analysis via STEM-EDS.

| Nominal chemical atomic ratio | Atomic ratio obtained from EDS spectra |
|---|---|
| Ru : Rh : Pd : Ir : Pt : Sb | Ru : Rh : Pd : Ir : Pt : Sb |
| 11.25 : 11.25 : 11.25 : 11.25 : 5 : 50 | 13.1 : 11.9 : 10.0 : 9.9 : 5.1 : 50.1 |
| 10 : 10 : 10 : 10 : 10 : 50 | 10.8 : 12.5 : 9.4 : 9.3 : 8.4 : 49.6 |
| 8.75 : 8.75 : 8.75 : 8.75 : 15 : 50 | 11.0 : 9.4 : 7.4 : 7.7 : 14.5 : 49.9 |
| 7.5 : 7.5 : 7.5 : 7.5 : 20 : 50 | 10.4 : 8.3 : 5.8 : 7.1 : 19.6 : 48.9 |

## c. $x$ dependence of unit call parameters for $M_{1-x}Pt_xSb$ (M = equimolar Ru, Rh, Pd, Ir)

The unit cell parameters of single-phase powder samples of $M_{1-x}Pt_xSb$ (M = equimolar Ru, Rh, Pd, Ir) were obtained from powder X-ray diffraction (XRD) patterns measured at room temperature with Cu-K$\alpha$ radiation using a diffractometer (Rigaku, Minflex). The analysis of the XRD patterns was performed by using SmartLab Studio II software (Rigaku). The XRD patterns and fitting curves for single-phase samples of $M_{1-x}Pt_xSb$ (M = equimolar Ru, Rh, Pd, Ir) with $x$ = 0.1, 0.2, 0.3 and 0.4 are shown in Fig. S2a. The excellent agreement was achieved by assuming NiAs-type structure. The obtained unit cell parameters by the fittings are summarized in Fig. S2b. Both the $a$ and $c$ show systematic dependence on $x$.

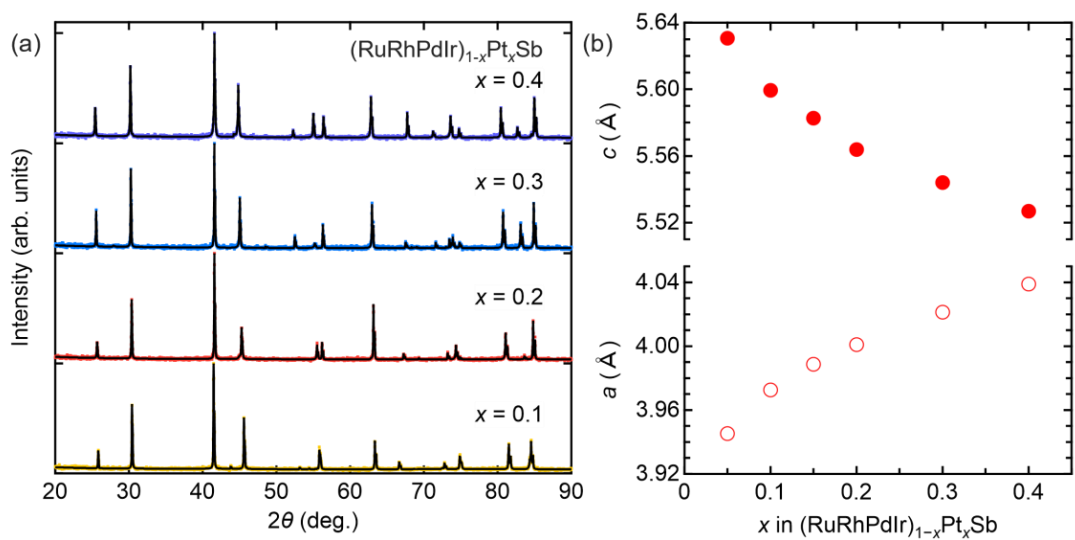

Fig. S2. (a) XRD patterns (dots) and fitting curves (solid lines) for single-phase samples of $M_{1-x}Pt_xSb$ (M = equimolar Ru, Rh, Pd, Ir) with $x$ = 0.1, 0.2, 0.3 and 0.4. (b) $x$ dependence of unit cell parameters for $M_{1-x}Pt_xSb$ (M = equimolar Ru, Rh, Pd, Ir) obtained from the profile fits.